\pdfoutput=1
\documentclass[conference]{IEEEtran}

\PassOptionsToPackage{hyphens}{url}
\usepackage[bookmarks=false]{hyperref}
\usepackage[utf8]{inputenc}
\usepackage{graphicx}
\usepackage{xcolor}
\usepackage{multirow}
\usepackage{makecell}
\usepackage{array}
\usepackage{amsmath,amssymb,amsfonts}
\usepackage{listings}
\usepackage{balance}
\usepackage[ruled,vlined]{algorithm2e}

\usepackage{pifont}
\usepackage{tikz}

\newcommand{\ignore}[1]{}

\usepackage[printwatermark]{xwatermark}
\newwatermark[firstpage,color=black!100,angle=0,scale=1,xpos=0,ypos=129]{\small\textnormal{© 2020 IEEE. Personal use of this material is permitted.  Permission from IEEE must be obtained for all other uses, in any current or \\ future media, including reprinting/republishing this material for advertising or promotional purposes, creating new collective works,  \\ for resale or redistribution to servers or lists, or reuse of any copyrighted component  of this work in other works.}}
\newwatermark[allpages,color=black!100,angle=0,scale=1,xpos=0,ypos=-126]{\small 2020 50th Annual IEEE/IFIP International Conference on Dependable Systems and Networks - Supplemental Volume (DSN-S) \\ Accepted version © 2020 IEEE. \url{https://doi.org/10.1109/DSN-S50200.2020.00023}}

\title{Secure Consensus Generation with Distributed DoH}

\author{\IEEEauthorblockN{Philipp Jeitner\IEEEauthorrefmark{1},
Haya Shulman\IEEEauthorrefmark{2} and Michael Waidner\IEEEauthorrefmark{1}\IEEEauthorrefmark{2}}
\IEEEauthorblockA{\IEEEauthorrefmark{1}Technical University of Darmstadt, philipp.jeitner@sit.tu-darmstadt.de}
\IEEEauthorblockA{\IEEEauthorrefmark{2}Fraunhofer Institute for Secure Information Technology SIT, \{haya.shulman,michael.waidner\}@sit.fraunhofer.de}
}

\date{March 2020}
\begin{document}

\maketitle

\begin{abstract}
Many applications and protocols depend on the ability to generate a pool of servers to conduct majority-based consensus mechanisms and often this is done by doing plain DNS queries. A recent off-path attack \cite{ntp-over-dns} against NTP and security enhanced NTP with Chronos \cite{ntp:chronos} showed that relying on DNS for generating the pool of NTP servers introduces a weak link. In this work, we propose a secure, backward-compatible address pool generation method using distributed DNS-over-HTTPS (DoH) resolvers which is aimed to prevent such attacks against server pool generation.
\end{abstract}

\section{Introduction}
Majority-based consensus mechanisms ensure security by comparing the responses from a set of servers taken from a larger pool. The security guarantee is given by statistical analysis and assumptions on the distribution of benign and malicious servers in that pool. To generate the pool, some applications rely on lists of servers signed by central authorities (e.g., Tor \cite{tor-directory-protocol}) but others like Chronos \cite{ntp:chronos} and most Cryptocurrencies \cite{dns-seed} just rely on the DNS. Recently \cite{cns:frag:dns,brandt2018domain,ntp-over-dns} showed that this reliance on DNS introduces a weak link and allows an attacker to attack DNS instead of the application protocol. Specifically \cite{ntp-over-dns} showed that Chronos is vulnerable to off-path attacks, by subverting DNS security, while Chronos \cite{ntp:chronos} guarantees security against strong on-path attackers. 
In this work we propose an approach for enhancing the DNS layer by distributing the queries for Network Time Protocol (NTP) servers to a set of resolvers. Our proposal, in tandem with Chronos, guarantees security to the NTP ecosystem. 

Previous proposals on distributing DNS \cite{cachin2004secure,swildens2010domain}, focused on the DNS nameservers, our goal in this work is to distribute the DNS resolvers. We propose a majority based protocol to securely generate a pool of servers via DNS which does not require any changes to the existing protocols nor infrastructure. Our proposal leverages majority decision supported by secure channels to a list of trusted DNS-over-HTTPS (DoH) \cite{RFC8484} resolvers. The security is guaranteed since distributing the resolvers ensures that the queries are sent from different locations in the Internet. Hence even if some of the paths or servers are corrupted or controlled by the attacker, the requests from the paths that are not under attacker's control guarantee security. The security of our proposal assumes that the realistic attackers are limited in their capabilities: the attacker can control some of the servers and some of the links in the Internet but not all the Internet servers and links. This is a realistic assumption and applies even to the strong government sponsored attackers.

\begin{figure}[ht!]
    \centering
    \vspace{-10pt}
    \includegraphics[width=0.35\textwidth]{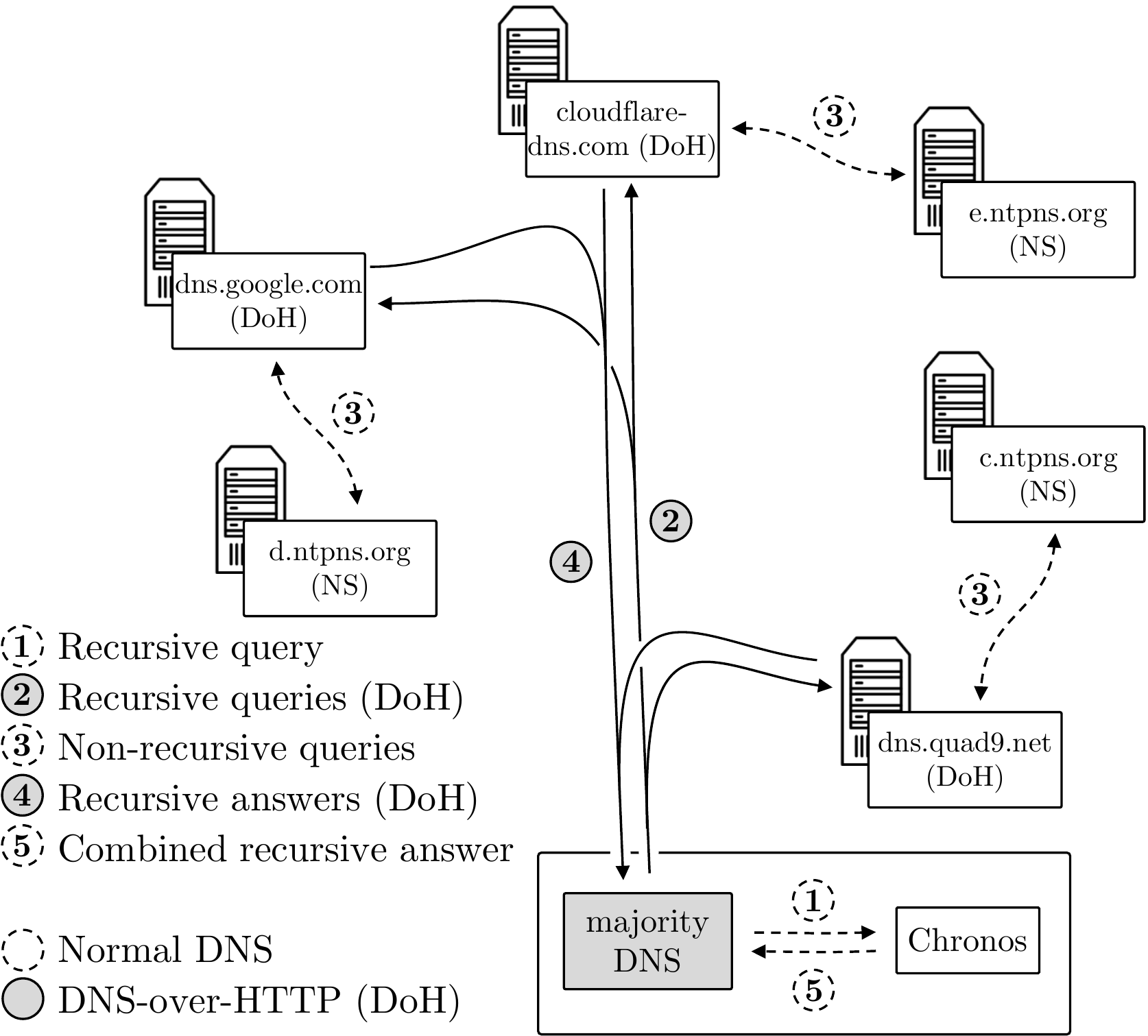}
    \vspace{-5pt}
    \caption{System overview}
    \vspace{-10pt}
    \label{fig:overview}
\end{figure}

 Our proposed system is shown in Figure~\ref{fig:overview}. The mechanism works by querying the NTP pool domain through a list of distributed DoH resolvers (Step 2 in Figure~\ref{fig:overview}) and combines the results to generate a server address pool containing a fraction of at least $x$ (e.g. $\frac{1}{2}$) benign servers assuming that an attacker can only attack a fraction of $x$ of the DoH resolvers. %

\section{Secure Server Pool Generation}

The operation mode targeted in this work is that of applications aiming to generate a trustworthy pool of servers, the application can connect to. For the application to be secure, this pool must include a fraction of at least $x$ benign servers. This is the operation mode Chronos \cite{ntp:chronos} (and to some degree, traditional Network Time Protocol (NTP) clients) operate in. In this operation mode, the application is aware that not all servers in the generated pool are trustworthy, but the number of these is limited to be lower than the fraction of $1-x$ of the total servers which addresses are included in the DNS response. As this operation mode is specific to server pool generation, it does only support address lookups\footnote{If Dual-stack operation needs to be supported, it depends on the application, whether the property of a honest majority of servers needs to be fulfilled for the union of A and AAAA records or for both sets individually.}. Assuming a fraction $x$ of secure DoH resolvers, this property can be fulfilled by querying all DoH resolvers, truncating the list of results from each resolver to the number of results from the resolver with the shortest list and then returning the the combination of these truncated lists to the application. This process is shown in Algorithm~\ref{algo:poolgen}.

\begin{algorithm}[t]
\footnotesize
\caption{Secure server pool generation lookup}
\label{algo:poolgen}
\SetAlgoLined
{\bf Input:} domain {\it (query domain)} \\
{\bf Input:} resolvers {\it (list of DoH resolvers)} \\
{\bf Input:} $x$ {\it (Fraction of assumed non-attacked resolvers, eg. $\frac{1}{2}$)} \\
results = [] , lengths = [] , addresspool = []\\
\For {res in resolvers} {
    r = query(res, domain) \\
    results.append(r) \\
    lengths.append($\|r\|$) \\
}
truncatelength = min(lengths[) \\
\For {r in results} {
    addresspool.add(truncate(r, truncatelength)) \\
}
{\bf return} addresspool
\end{algorithm}

Ensuring that \emph{all} of the servers in a returned DNS query are benign can be performed via a classic majority-vote on each of the returned addresses, e.g., the majority DNS resolver only includes an address in the final response, if it is given by a majority of the DoH resolvers. This is not required for Chronos since it can handle a minority of malicious servers by itself. %

We propose to deploy our mechanism without changing the DNS infrastructure, offering a standard-compatible DNS-resolver interface. The design is in Algorithm~\ref{algo:poolgen}. %

\section{Security Analysis}
In this section we give an analysis of the probability that responses from a fraction of $x$ DoH resolvers can be trusted assuming that (1) all resolvers are benign and (2) a probability that an attacker can successfully perform an attack against any of the resolvers independently with a probability of $p_\text{attack}$.

\paragraph{Fraction of DoH resolvers $x$ required for a successful attack}

For a successful attack against the DNS-using application, we assume the attacker needs to control at least $y$ (e.g., $\frac{1}{2}$) of the addresses in the server pool used by that application. Assuming a number of used DoH resolvers $N$ and the length of the shortest\footnote{We use the shortest list, because this prevents attacks where the attacker seeks to overwhelm resolvers by including more responses than usual (See attack against Chronos \cite{ntp-over-dns}). This comes at the cost of allowing DoS attacks when the attacker includes no responses at all in his poisonous response.} response list $K$, we get a total number of $N\cdot{}K$ addresses, where every resolver controls only $K$. To control more than $y$ addresses the attacker therefore needs to attack at least $x$ DoH resolvers fulfilling $yK \leq xK$, which gives us $x \ge y$. The attacker therefore is required to successfully attack at least $x = y$ of the DoH resolvers used.

\paragraph{Probability of attacking at least $x$ of DoH resolvers}

Assuming a number of used DoH resolvers $N$, the probability of attacking a fraction of at least $x$ is $p_\text{attack}^M$ with $M \leq \lceil x N \rceil$. Even when the only 3 DoH resolvers are used, this means that the probability of a successful attack which requires a malicious majority ($x \geq \frac{2}{3}$) is reduced significantly ($p_\text{attack}^2$). Furthermore, by increasing the number of used DoH resolvers, a successful attack becomes exponentially less probable, effectively giving the same type of asymptotic advantage over an attacker which is achieved by increasing the key size in a traditional cryptosystem.

\section{Discussion and Future Work}
For our approach to work, the application must handle multiple instances of the same address in the response as individual servers. Without this assumption, an attacker might overwhelm a majority of DoH resolvers if those resolvers do not give mutually exclusive sets of responses. Furthermore, our proposal is only meant to provide security on the DNS layer, namely we guarantee that even if some of the responses from the DNS resolvers can be attacked, the majority response is correct, i.e., the NTP servers returned in a response are benign.  This may however not always be the case: attackers can try to join the NTP pool themselves and operate malicious NTP servers. Hence, for the overall NTP ecosystem to maintain security a distributed mechanism on the NTP layer should also be used, such as the Chronos proposal \cite{ntp:chronos}.

\section{Conclusion}
In this work we propose a distributed generation of NTP server pool using multiple DoH DNS resolvers. We show that using our proposal mitigates the off-path attacks against plain NTP as well as against Chronos enhanced NTP \cite{ntp-over-dns}, and guarantees security even against realistic on-path Man-in-the-Middle (MitM) attackers that control some (but not all) of the Internet paths and DNS resolvers. Our proposal is easy to integrate and does not require changes to the DNS infrastructure, it is backward compatible with standard DNS nameservers. %

\bibliographystyle{IEEEtran}
\bibliography{main}

\begin{thebibliography}{1}
\providecommand{\url}[1]{#1}
\csname url@samestyle\endcsname
\providecommand{\newblock}{\relax}
\providecommand{\bibinfo}[2]{#2}
\providecommand{\BIBentrySTDinterwordspacing}{\spaceskip=0pt\relax}
\providecommand{\BIBentryALTinterwordstretchfactor}{4}
\providecommand{\BIBentryALTinterwordspacing}{\spaceskip=\fontdimen2\font plus
\BIBentryALTinterwordstretchfactor\fontdimen3\font minus
  \fontdimen4\font\relax}
\providecommand{\BIBforeignlanguage}[2]{{%
\expandafter\ifx\csname l@#1\endcsname\relax
\typeout{** WARNING: IEEEtran.bst: No hyphenation pattern has been}%
\typeout{** loaded for the language `#1'. Using the pattern for}%
\typeout{** the default language instead.}%
\else
\language=\csname l@#1\endcsname
\fi
#2}}
\providecommand{\BIBdecl}{\relax}
\BIBdecl

\bibitem{ntp-over-dns}
P.~{Jeitner}, H.~{Shulman}, and M.~{Waidner}, ``{The} {Impact} of {DNS}
  {Insecurity} on {Time},'' in \emph{2020 50th Annual IEEE/IFIP International
  Conference on Dependable Systems and Networks (DSN)}, 2020, (Forthcoming).

\bibitem{ntp:chronos}
O.~Deutsch, N.~R. Schiff, D.~Dolev, and M.~Schapira, ``Preventing ({Network})
  {Time} {Travel} with {Chronos},'' in \emph{Proceedings 2018 {Network} and
  {Distributed} {System} {Security} {Symposium}}, San Diego, CA, 2018.

\bibitem{tor-directory-protocol}
{The Tor Project}. Tor directory protocol, version 3.
  \url{https://github.com/torproject/torspec/blob/master/dir-spec.txt},
  accessed 2020-02-10.

\bibitem{dns-seed}
\BIBentryALTinterwordspacing
A.~F. Loe and E.~A. Quaglia, ``You shall not join: A measurement study of
  cryptocurrency peer-to-peer bootstrapping techniques,'' in \emph{Proceedings
  of the 2019 ACM SIGSAC Conference on Computer and Communications Security},
  ser. CCS ’19.\hskip 1em plus 0.5em minus 0.4em\relax New York, NY, USA:
  Association for Computing Machinery, 2019, p. 2231–2247. [Online].
  Available: \url{https://doi.org/10.1145/3319535.3345649}
\BIBentrySTDinterwordspacing

\bibitem{cns:frag:dns}
A.~Herzberg and H.~Shulman, ``Fragmentation {C}onsidered {P}oisonous: or
  one-domain-to-rule-them-all.org,'' in \emph{IEEE CNS 2013. The Conference on
  Communications and Network Security, Washington, D.C., U.S.}\hskip 1em plus
  0.5em minus 0.4em\relax IEEE, 2013.

\bibitem{brandt2018domain}
M.~Brandt, T.~Dai, A.~Klein, H.~Shulman, and M.~Waidner, ``{Domain Validation++
  For MitM-Resilient PKI},'' in \emph{Proceedings of the 2018 ACM SIGSAC
  Conference on Computer and Communications Security}.\hskip 1em plus 0.5em
  minus 0.4em\relax ACM, 2018, pp. 2060--2076.

\bibitem{cachin2004secure}
C.~Cachin and A.~Samar, ``Secure distributed dns,'' in \emph{International
  Conference on Dependable Systems and Networks, 2004}.\hskip 1em plus 0.5em
  minus 0.4em\relax IEEE, 2004, pp. 423--432.

\bibitem{swildens2010domain}
E.~S.-J. Swildens, R.~D. Day \emph{et~al.}, ``Domain name resolution using a
  distributed dns network,'' May~25 2010, uS Patent 7,725,602.

\bibitem{RFC8484}
P.~Hoffman and P.~McManus, ``Dns queries over https (doh),'' Internet Requests
  for Comments, RFC Editor, RFC 8484, October 2018.

\end{thebibliography}

\end{document}